# Multilayer Pt/Al Based Ohmic contacts for AlGaN/GaN Heterostructures Stable up to 600°C Ambient Air

Nitin Goyal[1,a], Srujana Dusari[1], Jochen Bardong[1], Farid Medjdoub[2], Andreas Kenda[1] and Alfred Binder[1]

[1]*Carinthian Tech Research CTR AG, Europastraße 4/1, Technologiepark Villach, A- 9524 Villach/ St. Magdalen, Austria*

[2]*Institute of Electronic, Microelectronic and Nanotechnology, Av. Poincaré, 59650 Villeneuve d'Ascq, France*

In this paper, we present a Pt/Al multilayer stack-based ohmic contact metallization for AlGaN/GaN heterostructures. CTLM structures were fabricated to assess the electrical properties of the proposed metallization. The fabricated stack shows excellent stability after more than 100 hours of continuous aging at 600°C in air. Measured I-V characteristics of the fabricated samples show excellent linearity after the aging. The Pt/Al-based metallization shows great potential for future device and sensor applications in extreme environment conditions.

## I. INTRODUCTION

Developments in high quality layer growth and device technology have led to high performance electronics based on GaN and SiC. Particularly, AlGaN/GaN based heterostructures have shown great potential for future high voltage switching as well as high-power RF applications. This is attributed to their superior physical and electrical properties[1,2]. The wide band gap of III-V nitride materials enable device operations at much higher temperatures than the conventional silicon based electronics. This can lead to their use in automotive, space and other harsh environment industrial applications. The core characteristics of GaN based high electron mobility transistor (HEMT) devices are the presence of temperature stable, highly mobile two-dimensional electron gas (2DEG) present at the AlGaN/GaN hetero-interface[1,2]. A temperature stable 2DEG in combination with a wide band gap enables AlGaN/GaN heterostructures to become an ideal technology platform for development of electronic devices for extreme environment applications. Different research groups have experimentally analyzed the material stability and electrical characteristics of AlGaN/GaN heterostructures up to 900°C[3-9]. It was observed that in-situ deposited SiN enables stability of GaN heterostructures up to 900°C[5-6]. However, much research is still needed to improve the device epistack structure as well as on the growth of transition layers between the GaN buffer and the substrate (Si, SiC) for improved device performance at high temperatures.

---

[a] Electronic mail: nitin.goyal@ctr.at

Apart from the growth, the most important system requirement is the availability of high-temperature stable electrical contacts. Ohmic contacts are among the most important parts of any device as they define the interface at the signal input and output. Unstable contacts can limit the device operations at high temperatures, especially in applications where replacement of components is extremely costly and/or is subject to delays. Furthermore, the degradation in ohmic contacts at high temperatures is irreversible and, therefore, thermally stable ohmic contacts are very crucial for high-temperature electronics. In fact, most electronic devices suffer loss of performance due to degradation in the metal contacts at high temperatures. Harsh environment applications demand electrical stability in terms of lifetimes of thousands of hours. Thus, stable and reliable contacts are among the most desirable features of high-temperature electronics. This puts very stringent requirements on the device technology development, including the metallization. Therefore, to develop GaN as a high-temperature technology platform for future harsh environment applications, it is imperative to achieve a stable ohmic metallization. This also requires material epistacks that are stable at such conditions.

In the past, many attempts to develop high-temperature ohmic metallization for AlGaN/GaN heterostructures have been reported. Most of these methods focus on the development of different metal stacks as cap on top of a Ti/Al based layer[12-27]. Some of these methods utilize refractory materials such as molybdenum. Other attempts suffered from degradation at high temperatures due to inter-diffusion effects. *Selvanathan et al*[11] used Mo based metal stacks and performed long term aging tests at 500°C, 600°C, and 700°C in an $N_2$ ambient. The Ti/Al/Mo/Au stack was stable for 45 hours at 600 °C and 8.5 hours at 700°C. *Dong et al*[26] performed thermal aging experiments in a pure $N_2$ atmosphere, and these devices show stability for 100 hours when treated between 400°C-500°C, but degradation at a much lower time line was observed when operated at temperatures above 500°C. Recently, Hu *et. al*[25] performed thermal aging experiments for n-GaN ohmic contacts in air at 600°C with devices showing degradation after 10 hours. *Borysiewicz et al*[27] studied $Ti_2AlN$-based multilayer ohmic contacts on n-GaN and performed aging tests in a quartz tube furnace open to ambient air with temperatures up to 500°C inside the furnace and a process time of up to 100 hours. Their samples showed ohmic behavior post aging, but with significant increase in contact resistance.

In this paper, we present high-temperature stable ohmic contacts for AlGaN/GaN heterostructures based on a Pt/Al multilayer stack on top of a Ti/Al adhesion layer. The fabricated structures were adapted for characterization by the circular transmission line method (CTLM), which was applied at room temperature after 100 hours of continuous annealing in a quartz-tube furnace in air ambient. The measurements showed that these ohmic contacts retained excellent stability after the aging process.

## II. EXPERIMENT

Custom designed metal organic chemical vapor deposition (MOCVD) grown GaN on Si 4'' wafer was ordered from a commercial wafer supplier. The layer structure consists of a 10 nm in-situ SiN cap and a 20 nm $Al_{0.18}Ga_{0.82}N$ layer on top of a 2 μm thick III-nitride buffer stack grown on a silicon substrate. Fig.1 shows schematically the different layers of the heterostructure.

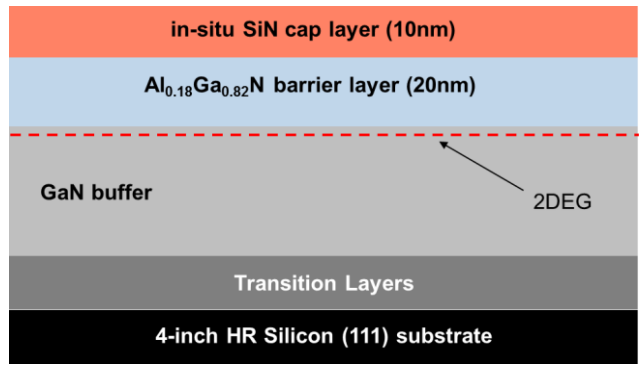

Fig.1 Schematic of the AlGaN/GaN heterostructure with different layers

CTLM structures are used to derive the sheet resistance and the contact resistance of the device. For this purpose, an optical mask was designed with four sets of CTLM structures on each die, with each set consisting of eight patterns. The CTLM patterns were designed with a constant outer radius R of 200 μm and gaps between the inner and outer radius varying with spacing *d* of 5, 10, 15, 20, 25, 30, 40, and 50 μm.

The CTLM structures were fabricated using conventional photolithography. Firstly, the GaN on Si wafer was diced into quarters. Prior to lithography and metallization, the samples were cleansed with acetone and isopropyl alcohol for 5 min each in an ultrasonic bath, then dipped in dilute HF: $H_2O$ (1:100) for 20 s and rinsed in deionized water. The samples were spin coated with photoresist and the CTLM layout was transferred to the resist by photolithography. After development of the resist, the SiN cap layer was etched by an optimized plasma reactive ion etching (RIE). Then, an in-situ surface pretreatment was performed using $SiCl_4$ plasma for 60 s in a RIE system at a chamber pressure of 25 mTorr.

Prior to loading the sample in the evaporation chamber, cleaning was performed in an HCl:H$_2$O solution to remove any surface oxide layer. The pressure in the evaporation chamber was maintained at 5x10$^{-8}$ Torr. The metallization stack deposited by evaporation consists of a patented layer pattern[28]: Ti (12nm)/Al (200nm)/ Pt (15nm) - Al (5nm)- Pt (15nm) - Al (5nm)- Pt (15nm) - Al (5nm)- Pt (15nm) - Al (5nm)- Pt (15nm) - Al (5nm) (Fig. 2). After deposition, the photoresist lift-off was performed in an acetone bath to leave the patterned metallization on the samples. Rapid thermal annealing was performed for 30 s at 875°C in an N$_2$ ambient. After that, sub-cell dicing was performed to obtain individual dies.

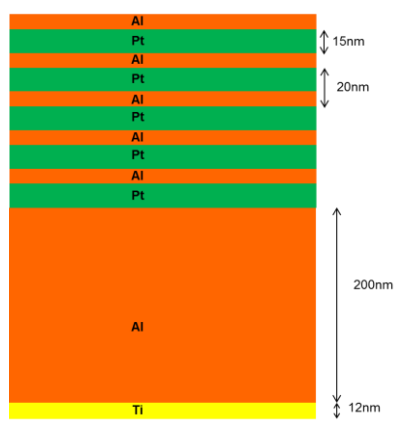

Fig.2 Schematic of the Ti/Al/multiple (Pt-Al) metal layer stack (layer thicknesses not to scale)

## III. RESULTS AND DISCUSSION

Device characterization was performed by using a manual wafer prober attached to a Keithley 2400 source meter and a corresponding LabVIEW program. I-V characteristics of the CTLM structures were automatically measured by applying voltage increments from -2.5V to +2.5V in 100 steps of 0.05V. Resistance values obtained using two probe method were corrected to account for unintentional error introduced due to the probes resistance. Before the aging of 100 hours, we measured the ohmic characteristics at room temperature of 32 CTLM test structures of each metallization, divided into 4 sets consisting of 8 structures each with varying circular gaps. After that, Pt/Al multi stack based test structures was exposed to a long-term aging test at 600°C for 100 hours using the previously mentioned quartz-tube furnace. As can be seen from Figure 3, the visual appearance of the Pt/Al stack based test structures remains almost unchanged by the aging process.

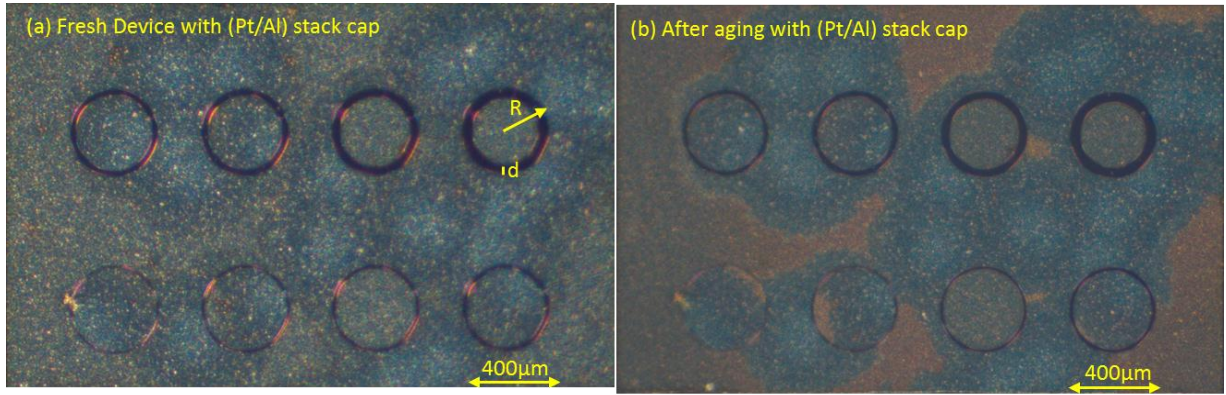

Fig.3 Fabricated CTLM structures based on Pt/Al multilayer stack with varying circular gaps between adjacent patterns as-formed (a) and after continuous aging of 100 hours at 600°C (b)

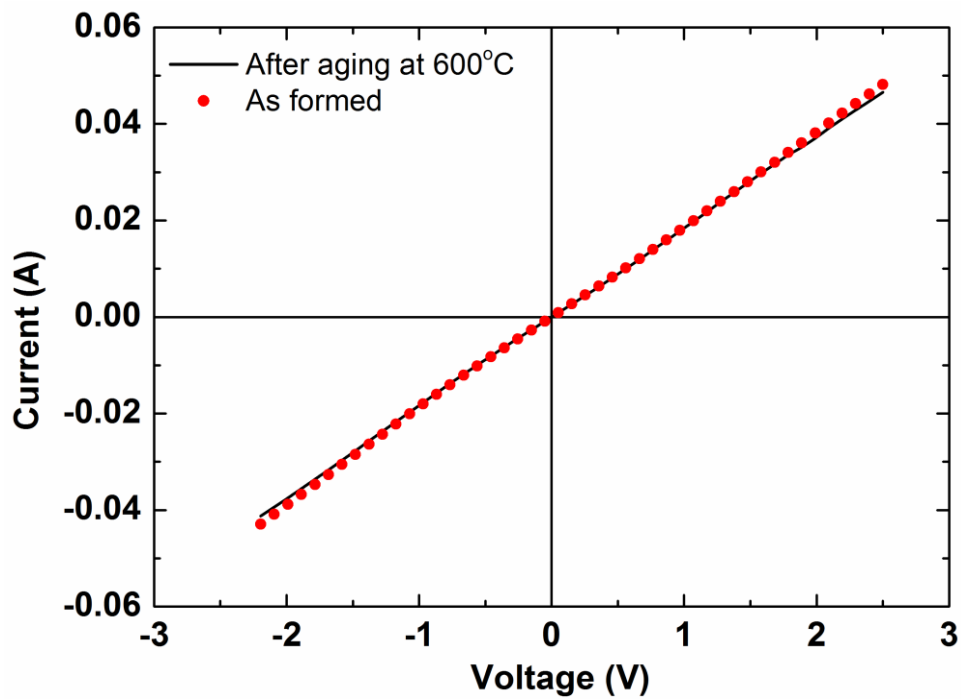

Fig.4 Comparison of measured I-V characteristics of the as-formed and an aged test structure (gap 20 µm) subjected to 100 hours of continuous aging at 600°C

After the aging process, different measurements were performed on the CTLM structures. First, the ohmic characteristics (I-V) were measured at room temperature. A comparison of the I-V characteristics of as-formed and aged devices based on Ti/Al/ (Pt/Al multilayer stack) is shown in Fig. 4.

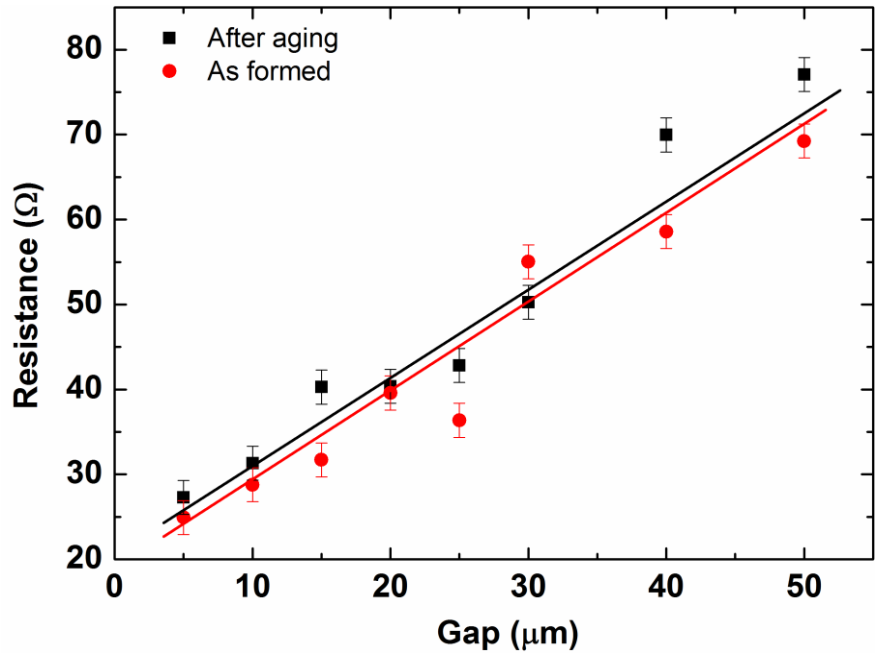

Fig.5 Comparison of CTLM characteristics of the as-formed and aged devices (Ti/Al/ (Pt/Al multilayer) stack) subjected to 100 hours of continuous aging at 600°C.

CTLM extracted resistances for two similar sets before and after aging are shown in Fig. 5. An interesting observation is that the contact resistance of the CTLM structures is reduced after the aging test at 600°C. This is due to the stabilization of Pt-Al stack towards low resistivity values at high temperatures, and the optimization of this is a matter of future investigations.

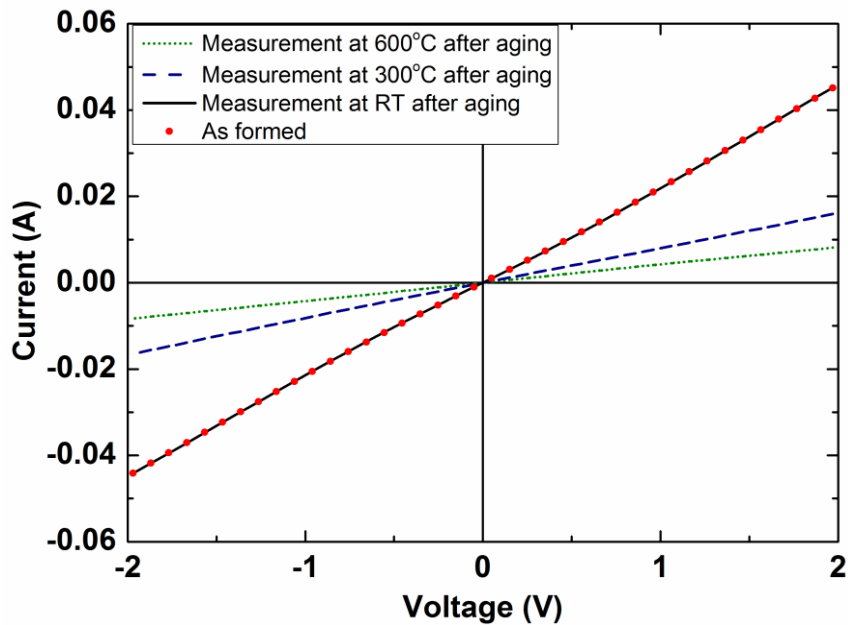

Fig.6 Comparison of I-V characteristics of the as-formed device, and devices measured at RT, 300°C, 600°C after aging

Subsequently, high temperature in-situ I-V measurements were performed using a hot chuck at a custom-built probe station with temperature set at 300°C and at 600°C (Fig.6). For all these measurements, custom made high temperature platinum probing tips were used. The measurements show that the I-V curves are linear and show a stable ohmic behavior when measured at high temperatures. The variation in current and slope is attributed to the change in electron mobility with temperature, which is a characteristic feature of all semiconductor devices (similar observation made by *Herfurth et. al.*[29]). Comparison of the ohmic characteristics between as-formed and aged test structures confirms the stability of the proposed metallization procedure. This shows that the multi-layer Pt-Al based metallization with a Ti/Al layer on AlGaN/GaN can be used for high-temperature ohmic contacts for GaN heterostructures. We attribute the stability of the Pt-Al multiple layer stacks at high temperature to the stable eutectic mixture that these thin metal films form when deposited in a ratio of 3:1 (Pt:Al). Contact resistivity values of as formed and aged test structures (treated at 600°C) are obtained using the CTLM method. Calculated values of contact resistivity before and after aging are $2.4 \times 10^{-7}$ Ω cm$^2$ and $2.5 \times 10^{-7}$ Ω cm$^2$ respectively. The resistance values show good thermal stability. The present metal stack offers very promising properties for applications requiring high temperature ohmic contacts, which are not available in other structures investigated to date[25, 30]. Presented metallization scheme can serve for GaN based sensing and device solutions that operate in extreme environment conditions.

At 600°C, not only the metal layers, but also the different material layers comprising the heterostructure expand and the lattice mismatch changes due to differences in the thermal expansion coefficients. Excessive mismatch in the epistack (on which metal layers are deposited) can result in the formation of dislocations and the reduction in the 2DEG, which, in turn, can lead to metallization failure. Carefully designed AlGaN and SiN layer stacks were used for this experiment, keeping in mind the buffer thickness of AlGaN for thermal expansion and strain relaxation at the elevated temperatures.

## IV. CONCLUSION

In conclusion, a stable ohmic metallization scheme has been developed for AlGaN/GaN heterostructures to sustain high-temperature environments. Linear I-V characteristics were obtained before and after aging. The contact resistance was found to be very stable after 100 hours of aging in air at 600°C. The stability achieved in the present devices shows a high degree of robustness of both the metallization scheme and the epitaxial structure design. This presents an opportunity and an important step towards the development of an AlGaN/GaN based high-temperature technology platform for device and sensor applications. The presented epi-stack design is well aligned with the conventional power device processes of AlGaN/GaN on Si.

## V. ACKNOWLEDGEMENT

This work is supported within the COMET – Competence centers for excellent technologies program by BMVIT, BMWFJ and the Austrian federal provinces of Carinthia and Styria. We acknowledge Prof. Johannes Sturm, FH Kärnten Villach, Austria for the timely equipment support. We also acknowledge Joff Derluyn, EpiGaN for useful discussions during this research work. We also acknowledge Prof. Tor A. Fjeldly, NTNU, Trondheim, Norway for the useful discussions and comments while preparation of this manuscript.

1. O. Ambacher, J. Smart, J. R. Shealy, N. G. Weimann, K. Chu, M. Murphy, W. J. Schaff, L. F. Eastman, R. Dimitrov, L. Wittmer, M. Stutzmann, W. Rieger, and J. Hilsenbeck, J. Appl. Phys. 85, 3222 (1999)

2. J. P. Ibbetson, P. T. Fini, K. D. Ness, S. P. DenBaars, J. S. Speck, and U. K. Mishra, Appl. Phys. Lett. 77, 250 (2000)

3. In Hak Lee, Yong Hyun Kim, Young Jun Chang, Jong Hoon Shin, T. Jang, Seung Yup Jang, Journal of the Korean Physical Society, Volume 66, Issue 1, pp 61-64 (2015)

4. J. Würfl, J. Hilsenbeck, E. Nebauer, G. Tränkle, and H. Obloh, Proc. GaAs Applications Symp. (GAAS), pp. 430-434, 1999

5. F. Medjdoub, J. Derluyn, K. Cheng, S. Degroote, M. Germain, G. Borghs, International Workshop on Nitride Semiconductors, Montreux, Switzerland, October 2008

6. M. Germain, J. Derluyn, M. Van Hove, F. Medjdoub, J. Das, D. Marcon, S. Degroote, K. Cheng, M. Leys, D. Visalli, P. Srivastava, K. Geens, J. Viaene, B. Sijmus, S. Decoutere and G. Borghs, CS MANTECH Conference, Portland, Oregon, USA, May 2010

7. N. Maeda, K. Tsubaki, T. Saitoh and N. Kobayashi, Appl. Phys. Lett. 79, 1634 (2001)

8. H. Tokuda, J. Yamazaki, and M. Kuzuhara, J. Appl. Phys. 108, 104509 (2010)

9. D. Donoval, M. Florovic, D. Gregušová, J. Kovác, P. Kordoš, Microelectronics Reliability 48, 1669–1672 (2008)

10. D. Selvanathan, L. Zhou, V. Kumar, and I. Adesida, phys. stat. sol. (a)194, No. 2, 583–586 (2002)

11. D. Selvanathan, F. M. Mohammed, A. Tesfayesus and I. Adesida, J. Vac. Sci. Technol. B 22, 2409 (2004)

12. L. Wang, F. M. Mohammed and I. Adesida, Appl. Phys. Lett. 87, 141915 (2005)

13. Ki Hong Kim, Chang Min Jeon, Sang Ho Oh, Jong-Lam Lee, and Chan Gyung Park, Jung Hee Lee, Kyu Seok Lee, Yang Mo Koo, J. Vac. Sci. Technol. B 23, 1 (2005)

14. F. M. Mohammed, L. Wang, D. Selvanathan, H. Hu and I. Adesida, J. Vac. Sci. Technol. B 23, 2330 (2005)

15. B. Van Daele, G. Van Tendeloo, W. Ruythooren, J. Derluyn, M. R. Leys and M. Germain, Appl. Phys. Lett. 87, 061905 (2005)

16. V. Desmaris, Jin-Yu Shiu, Chung-Yu Lu, N. Rorsman, H. Zirath, and Edward-Yi Chang, Journal of Applied Physics 100, 034904 (2006)

17. F. M. Mohammed, L.Wang, I. Adesida and E. Piner, J. Appl. Phys. 100, 023708 (2006)

18. H.T. Wang, L.S. Tan, E.F. Chor, Thin Solid Films 515, 4476–4479, (2007)

19. L. Wang, F. M. Mohammed, B. Ofuonye and I. Adesida, Appl. Phys. Lett. 91, 012113 (2007)

20. F. M. Mohammed, L. Wang, H. J. Koo and I. Adesida, J. Appl. Phys. 101, 033708 (2007)

21. L. Wang, F. M. Mohammed and I. Adesida, J. Appl. Phys. 101, 013702 (2007)

22. L. Wang, F. M. Mohammed, and Ilesanmi Adesida, J. Appl. Phys. 103, 093516 (2008)

23. D.A. Deen, D.F. Storm, D.S. Katzer, D.J. Meyer, S.C. Binari, Solid-State Electronics 54, 613–615, (2010)

24. S. S. Mahajan, A. Dhaul, R. Laishram, S. Kapoor, S. Vinayak, B.K. Sehgal, Materials Science and Engineering B 183, 47–53, (2014)

25. M. Hou, and D. G. Senesky, Appl. Phys. Lett. 105, 081905 (2014)

26. Z. Dong, J. Wang, C.P. Wen, S. Liu, R. Gong, M. Yu, Y. Hao, F. Xu, B. Shen, Y. Wang, Microelectronics Reliability 52, 434–438, (2012)

27. Michał A. Borysiewicz, Marcin Myśliwiec, Krystyna Gołaszewska, Rafał Jakieła, El zbieta Dynowska Eliana Kaminska, Anna Piotrowska, Solid-State Electronics 94, 15–19 (2014)

28. J. Bardong, G. Bruckner, R. Fachberger, B. Wall, "High temperature-resistant, Electrically Conductive Thin Films" United States Patent application publication "US 20130033150A1", Feb7, 2013

29. P. Herfurth, D. Maier, Y. Men, R. Rösch, L. Lugani, J.-F. Carlin, N. Grandjean and E. Kohn, *Semicond. Sci. Technol.* **28** 074026 (2013)

30. D. Maier, M. Alomari, E. Kohn, M. Diforte-Poisson, C. Dua, S.L. Delage, N. Grandjean, J.-F. Carlin, A. Chuvilin, U. Kaiser, D. Troadec, C. Gaquiere, 18th International Conference on Microwave Radar and Wireless Communications (MIKON), vol., no., pp.1,4, 14-16 June 2010